\documentclass[prl,aps,twocolumn,showpacs]{revtex4}
\usepackage{epsfig,amssymb,amsfonts,amsmath}
\def\id{\mathbb{I}}
\newcommand{\ket}[1]{|#1 \rangle}
\newcommand{\bra}[1]{\langle #1|}
\newcommand{\proj}[1]{\ket{#1}\bra{#1}}
\newcommand{\braket}[2]{\langle#1|#2\rangle}

\newcommand{\tr}[1]{\mbox{Tr}#1 }

\newcommand{\hilb}{\mathcal{H}}
\newcommand{\UU}{\mathbb{U}}

\begin{document}

\title{Fidelity and Coherence Measures from Interference}
%\date{\today}
%%%%%%%%%%%%%%%%%%%%%%%%%%%%%%%%%%%%%%%%%%%%%%%%%%%%%%%%%%%%%%%%%%
    \author{Daniel K. L. \surname{Oi}}
    %\email{D.K.L.Oi@damtp.cam.ac.uk}
    \affiliation{Centre for Quantum Computation, Department of Applied
      Mathematics and Theoretical Physics, University of Cambridge, Wilberforce
      Road, Cambridge CB3 0WA,
    United Kingdom}
    \author{Johan \surname{{\AA}berg}}
    \email{J.Aberg@damtp.cam.ac.uk}
    \affiliation{Centre for Quantum Computation, Department of Applied
      Mathematics and Theoretical Physics, University of Cambridge, Wilberforce
      Road, Cambridge CB3 0WA,
    United Kingdom}
%%%%%%%%%%%%%%%%%%%%%%%%%%%%%%%%%%%%%%%%%%%%%%%%%%%%%%%%%%%%%%%%%%

\begin{abstract}
  By utilizing single particle interferometry, the fidelity or coherence of a
  pair of quantum states is identified with their capacity for interference.
  We consider processes acting on the internal degree of freedom (e.g., spin or
  polarization) of the interfering particle, preparing it in states $\rho_{A}$
  or $\rho_{B}$ in the respective path of the interferometer. The maximal
  visibility depends on the choice of interferometer, as well as the locality
  or non-locality of the preparations, but otherwise depends only on the states
  $\rho_{A}$ and $\rho_{B}$ and not the individual preparation processes
  themselves. This allows us to define interferometric measures which probe
  locality and correlation properties of spatially or temporally separated
  processes, and can be used to differentiate between processes that cannot be
  distinguished by direct process tomography using only the internal state of the
  particle.
\end{abstract}

\pacs{03.67.-a}

\maketitle

A defining feature of quantum mechanics is the phenomenon of single particle
interference. The ability of a state to display interference, or of a quantum
process to preserve this ability, are intuitive notions of coherence and
coherent evolution. We elaborate this idea to define interferometric fidelity
and coherence measures, generalizing the coherent fidelities between quantum
channels introduced in Refs.~\cite{Oi2003,Aaberg2003}.  Interferometry has
played an important role in the development of theoretical concepts in quantum
mechanics, from which-way experiments~\cite{Englert1996} to geometric
phases~\cite{Pancharatnam1956}. By introducing mixed states and quantum
channels into this realm, we obtain a rich structure~\cite{Ann} which we
address in this Letter with focus on coherence. Apart from these fundamental
aspects, coherence is a prerequisite to quantum information processing.  To
construct practically useful fault tolerance and error correction schemes, it
is important to understand the coherence and correlation properties of the
processes that act in physical implementations~\cite{RPOR2005}.  The measures
put forward in this Letter provide means to probe these properties
interferometrically.

\begin{figure}
\includegraphics[width=0.45\textwidth]{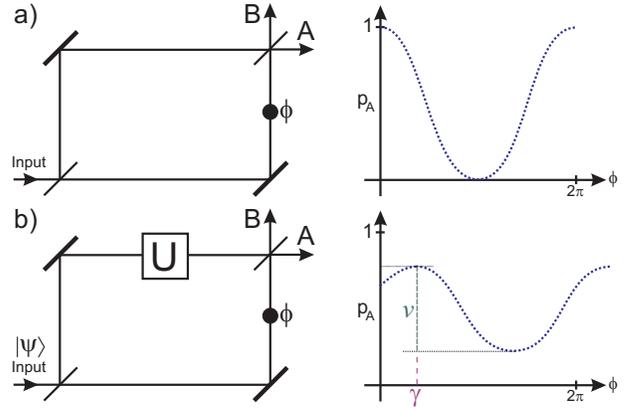}
\caption{a) Mach-Zender Interferometer. An initial beam splitter places an
  incident particle into a coherent superposition of traveling along the lower
  and upper paths. A phase shifter introduces a relative phase shift between
  the two paths before they recombine on a second beam-splitter, and the final
  direction of the particle is measured. b) A unitary operation in one path
  modifies the interference depending on the overlap between the interfering
  states $\ket{\psi}$ and $U\ket{\psi}$.}
\label{fig:machzender}
\end{figure}

We briefly review the Mach-Zender interferometer (Fig.~\ref{fig:machzender}a).
A single particle passes a beam splitter, which causes the particle to
traverse two paths in superposition allowing interference at a second
beam-splitter. The probabilities to detect the particle at the outputs,
$p_A=\frac{1}{2}(1+\cos\phi)$ and $p_B=1-p_A$, depend on a phase shift
$\phi$ in one path. The visibility
$v=[p_A(\phi_{max})-p_A(\phi_{min})]/[p_A(\phi_{max})+p_A(\phi_{min})]$ of the
interference pattern is unity when the two paths are perfectly coherent, and
there is no way, even in principle, to obtain any information about which path
the particle ``actually'' took~\cite{Englert1996}.

We now introduce an internal degree of freedom to the particle (e.g.
polarization, spin), described by a Hilbert space $\hilb_I$, and assume that
the beam-splitters and mirrors do not affect this internal state.  A unitary
operation $U$ acting on the internal state is placed in one path
(Fig.~\ref{fig:machzender}b).  If the internal state initially is
$\ket{\psi}$, this results in the new interference pattern $p_A =
\frac{1}{2}[1+v\cos(\phi-\gamma)]$ with visibility $v =
|\bra{\psi}U\ket{\psi}|$ and phase shift $\gamma=\arg(\bra{\psi}U\ket{\psi})$.
The internal state entangles with the path, and path information could be
extracted by using the distinguishability between $\ket{\psi}$ and
$U\ket{\psi}$ to a degree that corresponds to the reduction of visibility.
For a mixed input $\rho$, the interference can be expressed as $F(\rho)=v
e^{i\gamma}=\tr[\rho U]$~\footnote{We cannot directly interpret this as
  resulting from the distinguishability of the interfering states, but by
  considering purifications of the mixed states, we recover the
  interpretation.}.  We refer to $F$ as the \emph{interference function}. The
phase shift $\arg F(\rho)$ has been used to define parallel transport of mixed
states~\cite{SPEAEOV00,ESBOP02,faria02}, but here we consider the visibility
$|F(\rho)|$ of the interference effect.

\begin{figure}
 \includegraphics[width=0.45\textwidth]{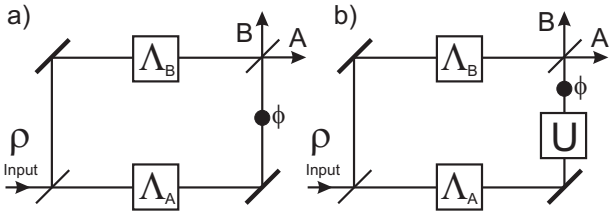}
 \caption{a) The internal state of the particle is affected by the channels
   $\Lambda_{A}$ or $\Lambda_{B}$. The gluing and the interference is not
   uniquely determined by $\Lambda_{A}$ and $\Lambda_{B}$.  b) A generalized
   Interferometer is obtained by inserting a variable unitary operator. This
   set-up can distinguish all gluings of two given channels.}
 \label{fig:mzcpmaps}
\end{figure}

If we insert into each path a process acting on the internal state of the
particle, it seems reasonable to ask how the interference is modified
(Fig.~\ref{fig:mzcpmaps}a). Suppose these processes can be described by
quantum channels, i.e., trace-preserving completely positive maps,
$\Lambda_{A}$ and $\Lambda_{B}$ respectively, what would be the corresponding
interference function?  Surprisingly~\cite{Oi2003,Aaberg2003,ESBOP02,faria02},
the interference is not determined solely by $\Lambda_{A}$ and $\Lambda_{B}$
but additional properties of the processes are required in order to uniquely
determine $F(\rho)$. More generally, the ``marginal'' channels $\Lambda_{A}$
and $\Lambda_{B}$ do not uniquely determine the joint operation $\Lambda$
acting on the two paths~\cite{Ann}. Adopting the terminology in
Refs.~\cite{Ann, Aaberg2003}, a joint channel $\Lambda$ is a \emph{gluing} of
the channels $\Lambda_{A}$ and $\Lambda_{B}$.

In many cases the two processes are independent, e.g., they occur at space-like
separation and do not pre-share either classical correlation or entanglement.
We call such a total operation a \emph{local subspace preserving} (LSP)
operation~\cite{Ann}.  It has been shown~\cite{Aaberg2003} that all possible
interference functions of LSP gluings of channels $\Lambda_{A}$ and
$\Lambda_{B}$ can be written as
\begin{equation}
\label{localfdef}
F(\rho) = \sum_{kl}b_{l}a_k^* \tr\left[A_{k}^{\dagger}B_{l}\rho\right],
\end{equation}
where $||\vec{a}||_2,\ ||\vec{b}||_2\leq 1$, and $\{A_{k}\}$ and $\{B_{l}\}$
are arbitrary but fixed linearly independent Kraus
representations~\cite{Kraus1983} of the channels $\Lambda_{A}$ and
$\Lambda_{B}$, respectively.

From Eq.~(\ref{localfdef}) one can see that it is possible to define an
operator $\widetilde{A}_{0} = \sum_{k}a_{k}A_{k}$, and similarly an operator
$\widetilde{B}_{0}$, such that $F(\rho) =
\tr[\widetilde{A}_{0}^{\dagger}\widetilde{B}_{0}\rho]$.  In the terminology of
Ref.~\cite{Oi2003} these are the coherence operators of the processes. By a
unitary transformation it is always possible to find a Kraus representation
with the coherence operator as one of the Kraus operators.  The coherence
operator then corresponds to the environment being undisturbed by the
particle, while the other operators represent cases when the environment
experiences a ``scattering event'' and the coherence of the particle is lost.

\begin{figure}
  \includegraphics[width=0.45\textwidth]{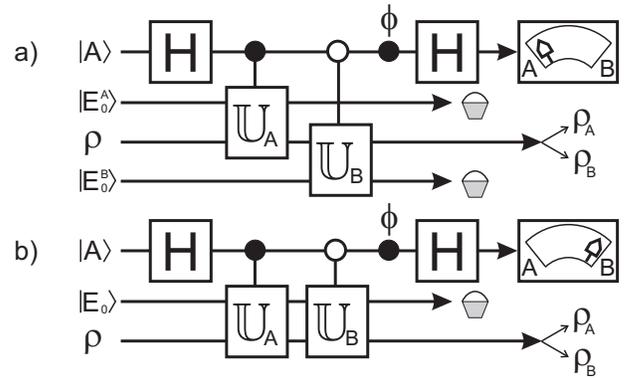}
  \caption{a) LSP gluing of $\Lambda_A$ and $\Lambda_B$ modeled by unitaries
    $\UU_A$ and $\UU_B$ acting on the internal state and separate ancillas. b)
    SP gluing of $\Lambda_A$ and $\Lambda_B$, where the paths share a common
    ancilla.}
\label{fig:cpmaps}
\end{figure}

Another approach to subspace local gluings is to use Stinespring
dilations~\cite{Stinespring1955} to represent the channel
$\Lambda_{A}(\rho)=\tr_{E_{A}}[\UU_{A}(\rho\otimes|E_{0}^{A}\rangle\langle
E_{0}^{A}|)\UU_{A}^\dagger]$, where $|E_{0}^{A}\rangle$ is a state of an
environment/ancilla.  Using a separate ancilla we can similarly represent
$\Lambda_{B}$.  It can be shown \cite{Aaberg2003} that all LSP Gluings of
$\Lambda_{A}$ and $\Lambda_{B}$ can be obtained as $\Lambda(\sigma) =
\tr_{E_{A}E_{B}}[\UU \sigma\otimes|E_{0}^{A}\rangle\langle E_{0}^{A}|\otimes
|E_{0}^{B}\rangle\langle E_{0}^{B}|\UU^{\dagger}]$, where $\UU =
|A\rangle\langle A|\otimes \UU_{A}\otimes\id _{B} + |B\rangle\langle B|\otimes
\id_{A}\otimes\UU_{B}$, by varying the Stinespring dilations
(Fig.~\ref{fig:cpmaps}a). Note that the coherence operators can be written
$\widetilde{A}_{0} = \langle E_{0}^{A}|\UU_{A}| E_{0}^{A}\rangle$ and
$\widetilde{B}_{0} = \langle E_{0}^{B}|\UU_{B}| E_{0}^{B}\rangle$, which
demonstrates that the choice of Stinespring dilations directly determine the
LSP gluing and hence the interference. One also sees that the coherence
operators indeed correspond to the case when the environment remains
unchanged, as mentioned above.
 
So far we have considered LSP gluings, but we may also consider more general
types of gluings. A \emph{subspace preserving} (SP) channel does not transfer
probability weight between the two paths, i.e. the particle does not ``jump'',
but apart from this restriction the process may use any communication or
shared classical or quantum correlations \cite{Ann}. In this case the
interference function is similar to Eq.~(\ref{localfdef}), but with
$b_{l}a^{*}_{k}$ generalized to a matrix $C_{lk}$ satisfying $CC^{\dagger}\leq
I$ \cite{Aaberg2003}. Similarly as for the LSP gluings, it can be shown
\cite{Aaberg2003} that all SP gluings can be reached through various choices
of Stinespring dilations of the glued channels, but with the difference that
the two paths share a common ancilla (Fig.~\ref{fig:cpmaps}b).

The ordinary interferometer has only a limited capacity to determine gluings.
By inserting a variable unitary operator in one arm (Fig.~\ref{fig:mzcpmaps}b)
we create a generalized interferometer whose generalized interference function
$G(\rho,U)$ can distinguish between all SP (and LSP) gluings of two given
channels~\cite{Aaberg2003},
\begin{equation}
\label{Gener}
G(\rho,U) = \sum_{kl}C_{lk}\tr\left[A_{k}^{\dagger}UB_{l}\rho\right].
\end{equation}

We can now define fidelity and coherence measures based on the maximum allowed
interference for given states, in analogy to Uhlmann's fidelity for
states~\cite{Uhlmann1976} (extended to channels in Ref.~\cite{Raginsky2001})
$\mathcal{F}_{Uhl}(\rho_A,\rho_A)=\sup_{\ket{\alpha},\ket{\beta}}\
|\braket{\alpha}{\beta}|$ where $\ket{\alpha}$ and $\ket{\beta}$ purify
$\rho_A$ and $\rho_B$, respectively.

If the particle initially is in the internal state $\ket{\psi}$, the first
beam-splitter causes the superposition
$\ket{\eta}=(\ket{A}\ket{\psi}+\ket{B}\ket{\psi})/\sqrt{2}$, where the
orthonormal states $\ket{A}$ and $\ket{B}$ correspond to the two paths of the
interferometer. We define the subspace local coherent fidelity
$\mathcal{F}^{(LSP)}(\rho_A,\rho_B)$ as the maximal visibility
achievable for all possible LSP operations preparing $\rho_{A}$
and $\rho_{B}$ in their respective path, i.e. all LSP operations $\Lambda$
such that $\bra{A}\Lambda(\proj{\eta})\ket{A}=\rho_{A}/2$ and
$\bra{B}\Lambda(\proj{\eta})\ket{B}=\rho_{B}/2$.  Hence,
\begin{equation}
\label{CohSL}
  \mathcal{F}^{(LSP)}(\rho_A,\rho_B)= \sup_{||\vec{a}||_2,||\vec{b}||_2\leq 1}
|F(\proj{\psi})|,
\end{equation}
where $\vec{a}$ and $\vec{b}$ are as in Eq.~(\ref{localfdef}). We might expect
that Eq.~(\ref{CohSL}) would depend on the choice of marginal channels
$\Lambda_{A}$ and $\Lambda_{B}$, and that we would have to optimize over all
channels such that $\Lambda_{A}(\proj{\psi})=\rho_{A}$ and
$\Lambda_{B}(\proj{\psi})=\rho_{B}$.  However, this is not the case as we
show below, $\mathcal{F}^{(LSP)}$ depends only on $\rho_{A}$ and
$\rho_{B}$, and we can choose any feasible channels $\Lambda_{A}$ and
$\Lambda_{B}$ to form the LSP gluings. Obviously, the choice of initial
internal state $\ket{\psi}$ does not matter, as long as it is pure.

Note that $F(\proj{\psi}) = \vec{a}^{\dagger}Q \vec{b}$,
where $Q$ is a matrix with elements
\begin{equation}
\label{Qdef}
Q_{kl} = \bra{\psi}A_{k}^{\dagger}B_{l}\ket{\psi}.
\end{equation}
It follows that the maximum of $|F(\proj{\psi})|$ (fixing $\Lambda_{A}$ and
$\Lambda_{B}$) is equal to the largest singular value of $Q$.

A set of (not necessarily normalized) pure states $\{\ket{a_{k}}\}$ is a
\emph{pure decomposition} of $\rho_{A}$ if $\rho_{A}=\sum_{k}\proj{a_k}$. Let
$\{\ket{b_{l}}\}$ be a pure decomposition of $\rho_{B}$ and consider the
matrix $M$ with elements $M_{kl} = \braket{a_{k}}{b_{l}}$. It can be shown
that the singular values of $M$ are independent of the choices of pure
decompositions. In particular, we may use the spectral decompositions with
eigenvalues $\lambda_k^{A(B)}$ and orthonormal eigenvectors
$|\widetilde{\psi}_k^{A(B)}\rangle$, to obtain $M_{kl}=(\lambda_k^A
\lambda_l^B)^{1/2}\braket{\widetilde{\psi}_k^{A}}{\widetilde{\psi}^{B}_l}$.
The $k^{\text{th}}$ singular value of $M$ is $s_{k}(M) =
\lambda_{k}(\sqrt{\sqrt{\rho_{B}}\rho_{A}\sqrt{\rho_{B}}})$, where
$\lambda_{k}$ is the $k^\text{th}$ eigenvalue of the enclosed operator.

Returning to Eq.~(\ref{Qdef}), since $\Lambda_{A}(\proj{\psi})=\rho_{A}$,
$\{A_{k}|\psi\rangle\}$ is a pure decomposition of $\rho_{A}$.  Similarly,
$\{B_{l}|\psi\rangle\}$ is a pure decomposition of $\rho_{B}$.  Thus, the
singular values of $Q$ are independent of the channels $\Lambda_{A}$ and
$\Lambda_{B}$ that generate $\rho_{A}$ and $\rho_{B}$, and we obtain
\begin{equation}
\mathcal{F}^{(LSP)}(\rho_A,\rho_B)=
\lambda_{max}\left(\sqrt{\sqrt{\rho_{B}}\rho_{A}\sqrt{\rho_{B}}}\right), 
\end{equation}
where $\lambda_{max}$ denotes the largest eigenvalue.

We define the subspace preserving coherent fidelity
$\mathcal{F}^{(SP)}(\rho_A,\rho_B)$ similarly as for
$\mathcal{F}^{(LSP)}$, but allowing all SP gluings. It can be shown that
the maximum of $\left|F\left(\proj{\psi}\right)\right|$ for all SP gluings is equal
  to $\sup_{C C^{\dagger}\leq I}\left|\tr\left[C Q\right]\right| =
  \sum_{k}s_{k}(Q)$, with $Q$ defined in Eq.~(\ref{Qdef}).  Since the singular
  values of $Q$ are independent of the chosen channels,
\begin{equation}
\mathcal{F}^{(SP)}(\rho_A,\rho_B)=
\tr\sqrt{\sqrt{\rho_{B}}\rho_{A}\sqrt{\rho_{B}}} = \mathcal{F}_{Uhl}(\rho_A,\rho_B).
\end{equation}
This result is also obtainable from the Stinespring construction of
the SP gluings.  For all purifications $\ket{\alpha}$ and $\ket{\beta}$ of
$\rho_A$ and $\rho_B$, there are Stinespring dilations such that the
resulting gluing (Fig.~\ref{fig:cpmaps}b) implements the transformation
$(\ket{A}+\ket{B})|\psi\rangle\ket{E_0}/\sqrt{2}\rightarrow
\left(\ket{A}\ket{\alpha}+\ket{B}\ket{\beta}\right)/\sqrt{2}$, which has
visibility $v=|\braket{\alpha}{\beta}|$, for which the maximum over all
purifications is the Uhlmann fidelity.

The coherent fidelities measure the coherent \emph{overlaps} of the two states
interfering at the beam-splitter. In keeping with the notion that unitary
operations preserve coherence (though not necessarily the fidelity) of states,
we would like to characterize purely the coherence of a preparation.  For
example, if $\ket{\psi}$ and $\ket{\psi^{\perp}}$ are orthogonal, a possible
global state is $(\ket{A}\ket{\psi}+\ket{B}\ket{\psi^{\perp}})/\sqrt{2}$, but
the coherent fidelity measures are zero. However, by a subspace local unitary
transformation rotating $\ket{\psi^{\perp}}$ into $\ket{\psi}$, we may regain
the maximal visibility reflecting this potential capacity for interference.
To quantify this, we employ the generalized interferometer and define
$\mathcal{G}^{(LSP)}(\rho_A,\rho_B)$ between two states $\rho_{A}$ and
$\rho_{B}$ as the maximal visibility that can be reached for all possible
unitary shifts $U$ and for all possible LSP operations that prepare the states
$\rho_{A}$ and $\rho_{B}$ (Fig.~\ref{fig:mzcpmaps}b).  If we initially fix
$U$, the calculation of is almost as for previous measures, except that
Eq.~(\ref{Qdef}) is replaced with
$\widetilde{Q}_{kl}=\bra{\psi}A_{k}^{\dagger}U B_{l}\ket{\psi}$.  We find that
the singular values of this matrix are independent of the choices of feasible
channels, yielding
\begin{eqnarray}
\mathcal{G}^{(LSP)}(\rho_A,\rho_B)&=&
\sup_{U}\lambda_{max}\left(\sqrt{\sqrt{\rho_{B}}U\rho_{A}U^{\dagger}\sqrt{\rho_{B}}}\right)
\nonumber\\
&=&
\sqrt{\lambda_{max}(\rho_{A})}\sqrt{\lambda_{max}(\rho_{B})}.
\end{eqnarray}
Note that this measure is a product between two quantities each related only
to local objects. Similarly, allowing SP operations in the generalized
interferometer yields
\begin{eqnarray}
\mathcal{G}^{(SP)}(\rho_A,\rho_B)&=&
\sup_{U}\tr\sqrt{\sqrt{\rho_{B}}U\rho_{A}U^{\dagger}\sqrt{\rho_{B}}}\nonumber\\
&=&
\sum_{k}\sqrt{\lambda_{k}^{\downarrow}(\rho_{A})}\sqrt{\lambda_{k}^{\downarrow}(\rho_{B})},
\end{eqnarray}
where $\lambda_{k}^{\downarrow}$ denotes the eigenvalues of the enclosed
operator sorted in a non-increasing order.

\begin{figure}
  \includegraphics[width=0.45\textwidth]{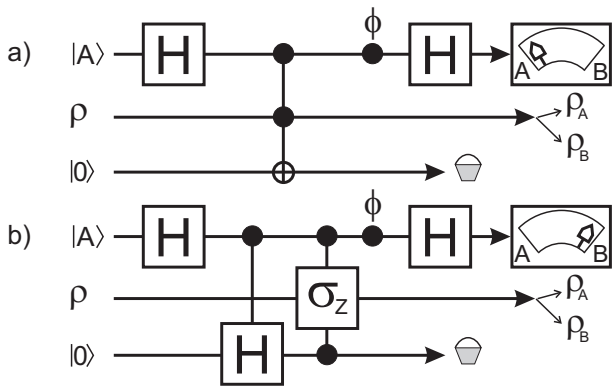}
  \caption{In the upper path of an interferometer is a channel that
    converts all input states into incoherent mixtures of the basis states,
    i.e., transverse relaxation (or $T_2$ process in spin dynamics). This
    can arise either from, e.g., a) an effective measurement modeled
    by a controlled-NOT onto a measurement qubit, b) by a Hadamard gate
    applied on the ancillary qubit, followed by a $\sigma_{z}$-gate
    conditioned on the ancilla.}
\label{fig:alternatenetworks}
\end{figure}

The interferometric measures introduced in this Letter lend themselves to
experiment, and can serve as tools to probe the dependence or independence of
physical processes.  Since these measures are independent of the marginal
channels ($\Lambda_{A}$ and $\Lambda_{B}$), this technique may be used even if
we do not know them. If the input internal state is pure it suffices to know
the states $\rho_{A}$ and $\rho_{B}$. If the visibility exceeds what is
obtainable with LSP gluings, then we can conclude that the total process
cannot be subspace local.  More generally, the interferometric approach allows
us to differ between processes, although these are indistinguishable if
regarded as operations on the internal state of the particle.  For example,
consider the transverse relaxation of a qubit, turning all input qubit states
into incoherent mixtures of the $\ket{0}$ and $\ket{1}$ states. This channel
can be obtained as the average output from a projective measurement of
$\sigma_Z$ (Fig.~\ref{fig:alternatenetworks}a). The maximum visibility
obtainable with this process in one path of an interferometer is $v_{max}=1$.
The same marginal channel results from the circuit outlined in
Fig.~\ref{fig:alternatenetworks}b, however the maximal visibility is
$v_{max}=1/\sqrt{2}$ in this case. Thus, the interferometer distinguishes
between these two processes, although a direct process tomography on the
internal state would reveal no difference.

Although we have described the interferometer in terms of spatially separated
paths and processes, these interferometric techniques can also be applied to
temporally separated processes. Time-bin photonic qubits, described in
Ref.~\cite{BGTZ1999}, propagating in a material with minimal effect on the
timing of the pulses, may serve as the spatial degree of freedom in
Fig.~\ref{fig:machzender}, while the polarization of the photons corresponds
to the internal degree of freedom.  If the polarization decoherence is due to
``classical'' perturbations of the optical fibre (e.g.  vibration or thermal
stresses) the visibility could be expected to be high for short time delays,
since the two pulses experience highly correlated noise and thus the gluing
would be described by an SP operation. If the delay time exceeds the
auto-correlation time of the polarization distortions, the gluing can be
described by an LSP operation and the visibility would drop below the
threshold for independent quantum channels~\cite{Oi2003}.

In conclusion, we define interferometric measures of fidelity and coherence
between states.  These quantify the ``quantumness'' of the preparation
processes, in the sense that they correspond to the capacity of the operations
to preserve the ability of the particles to interfere.  We define four measures
based on the maximal visibility obtainable in an interferometer, differing with
respect to the locality or non-locality of the preparation procedures, as well
as the choice of interferometer. In the case of the standard Mach-Zender
interferometer and non-local operations in the form of subspace preserving
channels~\cite{Ann} we obtain the Uhlmann fidelity as the maximal visibility.
The operational nature of these measures lend themselves to experiment, as well
as for investigating coherence and correlation properties of spatially or
temporally separated physical processes, which is important for the tuning of
fault tolerant and error correction schemes to the characteristics of the
underlying processes causing decoherence~\cite{RPOR2005}.

One can consider extending the ideas presented in this Letter, allowing mixed
input states, the concatenation of several channels which could be
generalized to (Markovian) continuous quantum channels, and analogous
measures for the channels themselves.

\begin{acknowledgments}
  DKLO acknowledges the support of the Cambridge-MIT Institute Quantum
  Information Initiative, EU grants RESQ (IST-2001-37559) and TOPQIP
  (IST-2001-39215), EPSRC QIP IRC (UK), and Sidney Sussex College, Cambridge.
  J{\AA} acknowledges support from the Swedish Research Council.
\end{acknowledgments}


\begin{thebibliography}{99}
\bibitem{Oi2003}D. K. L. Oi, Phys. Rev. Lett. \textbf{91}, 067902 (2003)
\bibitem{Aaberg2003}J. {\AA}berg, Phys. Rev. A \textbf{70}, 012103 (2004)
\bibitem{Englert1996}B-G Englert, Phys. Rev. Lett. \textbf{77}, 2154 (1996)
\bibitem{Pancharatnam1956} S. Pancharatnam, {\em Proc. Ind. Acad. Sci. A}
  \textbf{44}, 247 (1956).
\bibitem{Ann} J. {\AA}berg, Annals of Physics \textbf{313}, 326 (2004)
\bibitem{RPOR2005}
%Quantum-gate characterization in an extended Hilbert space 
P. P Rohde, G. J. Pryde, J. L. O'Brien, T. C. Ralph, Phys. Rev. A \textbf{3}, 032306 (2005)
\bibitem{SPEAEOV00}E. Sj\"oqvist, A.K. Pati, A. Ekert, J.S. Anandan, M.
  Ericsson, D.K.L. Oi, and V. Vedral, \emph{Phys. Rev. Lett.}  \textbf{85},
  2845 (2000). 
\bibitem{ESBOP02}M. Ericsson, E. Sj\"oqvist, J. Br\"annlund, D. K. L. Oi, A.
  K. Pati, Phys. Rev. A. \textbf{67}, 020101 (2003)
\bibitem{faria02}J. G. Peixoto de Faria, A. F. R. de Toledo Piza, M. C. Nemes,
%Phases of quantum states in completely positive non-unitary evolution.
Europhysics Letters. \textbf{62}, 782 (2003)
\bibitem{Kraus1983}K. Kraus, {\em States, Effects, and Operations},
   Springer-Verlag, Berlin (1983)
\bibitem{Stinespring1955} W. F. Stinespring,
%Positive Maps on C$^*$-Algebras,
Proc. Amer. Math. Soc. \textbf{6}, 211 (1955)
%% \bibitem{Jamiolkowski1972} A. Jamio{\l}kowski, \emph{Rep. Math. Phys.} {\bf 3}, p275 (1972)
\bibitem{Uhlmann1976}A. Uhlmann,
%% %The transition probability in the state space of a *-algebra,
{\em Rep. Math. Phys.} {\bf 9}, p273 (1976)
\bibitem{Raginsky2001}M. Raginsky,
%A Fidelity Measure for Quantum Channels,
{\em Phys. Lett. A} {\bf 290}, p11 (2001)
%% \bibitem{HJ1985}R. A. Horn, C. R. Johnson, {\em Matrix Analysis}, Cambridge
%%   University Press (1985)
\bibitem{BGTZ1999}J. Brendel, N. Gisin, W. Tittel, and H. Zbinden,
%Pulsed Energy-Time Entangled Twin-Photon Source for Quantum Communication,
Phys. Rev. Lett. \textbf{82}, 2594 (1999)

\end{thebibliography}
\end{document}